\documentclass[useAMS,usenatbib]{mn2e}

\usepackage[english]{babel}
\usepackage{amsfonts}
\usepackage{amsmath}
\usepackage{amssymb}
\usepackage{graphicx}
\usepackage{epstopdf}
\usepackage{times}
\usepackage{natbib}

\newif\ifAMStwofonts
\AMStwofontstrue

\begin{document}

\title{Lensing dispersion of supernova flux: a probe of nonlinear structure growth}

\author[C. Fedeli et al.]{C. Fedeli$^1$ and L. Moscardini$^{2,1,3}$\\
	$^1$ INAF - Osservatorio Astronomico di Bologna, Via Ranzani 1, 40127 Bologna, Italy\\
	$^2$ Dipartimento di Fisica e Astronomia, Universit\`a di Bologna, Viale Berti-Pichat 6/2, 40127 Bologna, Italy\\
	$^3$ INFN - Sezione di Bologna, Viale Berti-Pichat 6/2, 40127 Bologna, Italy\\}
	
\maketitle

\begin{abstract}
The scatter in the apparent magnitude of type Ia supernovae induced by stochastic gravitational lensing is highly dependent on the nonlinear growth of cosmological structure. In this paper, we show that such a dependence can potentially be employed to gain significant information about the mass clustering at small scales. While the mass clustering ultimately hinges on cosmology, here we demonstrate that, upon obtaining more precise observational measurements through future cosmological surveys, the lensing dispersion can very effectively be used to gain information on the poorly understood astrophysical aspects of structure formation, such as the clumpiness of dark matter halos and the importance of gas physics and star formation into shaping the large-scale structure. In order to illustrate this point we verify that even the tentative current measurements of the lensing dispersion performed on the Supernova Legacy Survey sample favor a scenario where virialized structures are somewhat less compact than predicted by $n-$body cosmological simulations. Moreover, we are also able to put lower limits on the slope of the concentration-mass relation. By artificially reducing the statistical observational error we argue that with forthcoming data the stochastic lensing dispersion will allow one to importantly improve constraints on the baryonic physics at work during the assembly of cosmological structure.
\end{abstract}

\section{Introduction}\label{sct:introduction}

One of the most powerful instruments for measuring the geometry of the Universe is given by standard candles: sources of light that are known to have all the same intrinsic luminosity, or at least whose intrinsic luminosity can be standardized. In spite of the many kinds of standard candles that have been proposed in recent years (gamma-ray bursts, active galactic nuclei, etc.), none can yet rival the accuracy achieved by type Ia supernovae (SN Ia henceforth). Indeed, SN Ia provided the first direct evidence of accelerated expansion of the Universe, thus exposing the need for dark energy \citep{RI98.1,PE99.1}. Despite this success, there are a number of factors contributing to the statistical uncertainties of supernova-based measurements, besides the ordinary measurement errors. First, the intrinsic luminosity of SN Ia is commonly standardized through the shape of their luminosity curves \citep{GU05.1,GU07.1}. This method has its own intrinsic uncertainties, mainly because the internal working of SN Ia is not yet fully understood, so that it is indeed not certain that all supernovae can be reduced to have the exact same intrinsic luminosity irrespective of environment, metallicity, and epoch. Second, the light coming from different supernovae is absorbed by different amounts of dust, either in the host galaxy and in the intergalactic space. Third, but not least, the light emitted by different SN Ia passes through different  sections of the large-scale structure (LSS), and hence it is differently affected by the gravitational deflection of light. Because the Universe is isotropic the average change in the apparent magnitude of a standard candle induced by this last effect should be vanishing. The dispersion, however, should not.

In this context, dust absorption can be distinguished from gravitational lensing because the former $i)$ is frequency dependent and $ii)$ produces only an apparent dimming, while the latter does not. Given this, if one has the intrinsic luminosity dispersion of SN Ia under control (or at least can parametrize it and marginalize over it), then he or she can deduce the lensing-induced dispersion \citep{ME99.1}. Alternatively, the same result can be obtained by modeling the magnification contribution given by the galaxies that happen to be close to the line of sight to the observed supernova \citep{ME01.1,JO10.1,KR10.1}. This lensing-induced dispersion (or briefly \emph{lensing dispersion}) is not just a mere source of noise, but rather contains extremely important information on the growth of cosmic structures all the way from the SN Ia redshift until now. Indeed, the lensing dispersion has been proposed as a tool for inferring cosmological parameters in a seminal work by \citet*{BE97.1}, then expanded by other authors \citep*{HA00.1,DO06.2,MA13.1,QU13.1}. In doing so however, a key factor has often been neglected: the lensing effect is highly sensitive to the matter clustering on small scales. This is indeed dependent on cosmology, particularly the normalization $\sigma_8$ of the mass power spectrum, nonetheless using the lensing dispersion for cosmological inference implies that we understand correctly the nature of dark matter \citep{ME99.2,ME07.1} and the growth of structures at large spatial frequencies.

The small-scale clustering of dark matter can be modeled only through numerical $n-$body simulations, because the mode coupling induced by the nonlinear growth of structures cannot be followed by linear perturbation theory. Inasmuch as cosmological simulations have difficulty to resolve small scales, significant uncertainties still persist on modeling the dark matter power spectrum for $k\gtrsim 30-50 h$ Mpc$^{-1}$ \citep{PE94.1,SM03.1}. Even more important, small scales are substantially affected by the physics of baryonic matter, namely gas and stars. The latter can also be studied through numerical (hydrodynamic) simulations, however these are much more time consuming than simple $n-$body runs, and their results depend heavily on the approximations involved in modeling baryonic physical processes (\citealt{JI06.1}; \citealt*{RU08.2}; \citealt{VA11.1}; \citealt*{FE12.1}). As a consequence, the clustering of baryons is still very much open to question, and the clustering of mass in general for $k\gtrsim 10h$ Mpc$^{-1}$ is still poorly understood. 

The immediate consequence of these considerations is that before using the lensing dispersion of SN Ia to constrain cosmology it is necessary to understand how the dispersion itself depends on the uncertainties of the nonlinear clustering of matter. Even more relevant, if this dependency is strong enough one can select a specific cosmology and use the observed lensing dispersion to constrain the inadequately understood small-scale structure growth. In this paper, we give a first assessment of these issues. First we study the reliance of the supernova lensing dispersion on the nonlinear matter power spectrum. Then, we use a simple model for the small-scale mass clustering in order to show that the even the marginal present measurements of the dispersion can provide interesting information on structure growth. Throughout the paper, where cosmology is needed we adopt a flat $\Lambda$CDM model based on the recent CMB results from the \emph{Planck} satellite \citep{PL13.1}: $\Omega_{\mathrm{m},0} = 0.3175$, $\Omega_{\mathrm{b},0} = 0.0490$, $H_0=h100$ km s$^{-1}$ Mpc$^{-1}$ with $h=0.6711$, $\sigma_8=0.8344$, and $n=0.9624$.

\section{Lensing-induced dispersion of supernova apparent magnitudes}\label{sct:scatter}

The scatter in the luminosity distance (and related quantities) induced by stochastic matter density fluctuations in the Universe via the gravitational deflection of light has been estimated in the past by using the relation between lensing convergence and magnification in the weak deflection regime \citep{ME99.1,BA01.1,DO06.2}. However, in a recent series of papers Ben-Dayan and collaborators \citep{BE13.2,BE13.1,BE13.3} computed the mean lensing dispersion $\sigma_\mu(z)$ of the distance modulus for a source placed at redshift $z$ using the full light-cone and ensemble averages in a relativistic setting\footnote{For modifications due to the peculiar motion of the observer we refer to the recent work by \citet{FA13.1}.}. This computation rests on several approximations, which the authors of \citet{BE13.2} estimated to affect the resulting dispersion at the level of $\sim 10\%$. Even larger uncertainties would however not be very important for the scope of the present work, given the still tentative observational detection of the lensing dispersion and the illustrative nature of this paper. Therefore we used this result, according to which the lensing dispersion depends on the spectrum of potential fluctuations via

\begin{eqnarray}\label{eqn:dispersion}
\sigma^2_\mu(z) &=& \frac{5\;\mathrm{mag}^2}{3}\;\mathrm{Log}^2(e) \int_0^{+\infty}\frac{\mathrm{d}k}{k} \Delta^2_\Psi(k,z) \times \nonumber\\
&\times& \left[ k\;c\Delta\eta(z) \right]^3\int_0^{k\;c\Delta\eta(z)} dx \frac{\sin x}{x}
\end{eqnarray}
\citep{BE13.2}. In the former Equation $\Delta\eta(z)$ is the difference in conformal time $\eta$ between today and redshift $z$. It can be easily written as

\begin{equation}
\Delta\eta(z) = \int_0^z \frac{\mathrm{d}z}{H(z)}\;.
\end{equation}
The function $\Delta^2_\Psi(k,z)$ represents the dimensionless power of the peculiar LSS potential, which can be recast in terms of the matter power spectrum $P(k,z)$ as (e.g., \citealt{ME99.1})

\begin{equation}
\Delta^2_\Psi(k,z) = \frac{9}{8\pi^2} \frac{H_0^4\Omega_{\mathrm{m},0}^2}{c^4k}(1+z)^2P(k,z)\;.
\end{equation}
We verified that the lensing dispersion computed according to the expression found in past papers (see for instance Eq. (5) of \citealt{DO06.2}) differs from its counterpart computed via Eq. (\ref{eqn:dispersion}) by at most $\sim 15\%$ (at $z\sim 2$). For the same reasons explained above, this difference is unimportant for the scope of the present paper.

\begin{figure}
	\centering
	\includegraphics[width=\hsize]{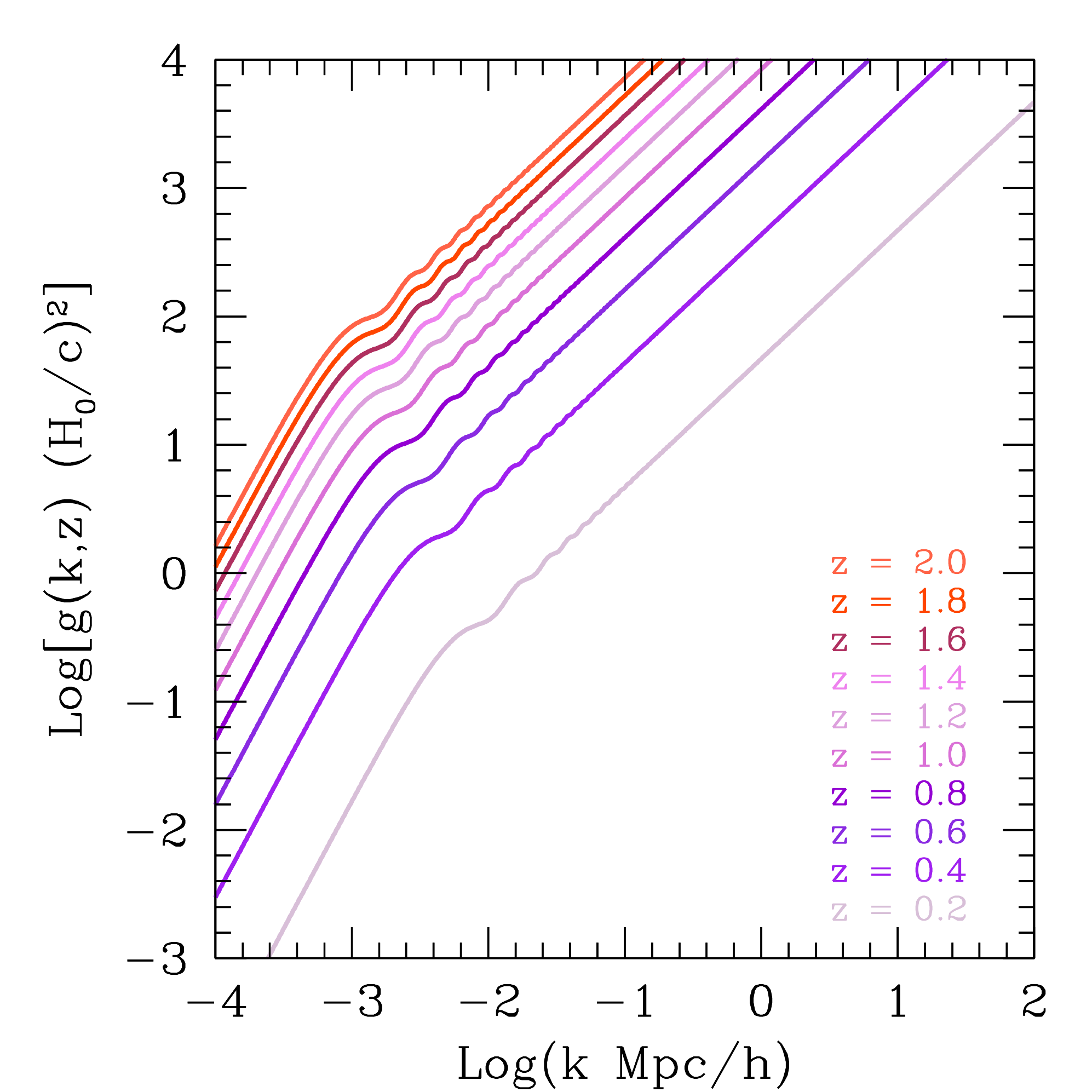}
	\caption{The integration kernel $g(k,z)$ for the dispersion in the distance modulus, defined in Eq. (\ref{eqn:kernel}), as a function of spatial frequency $k$. Results are shown for ten different redshifts, equally spaced between $z=0.2$ and $z=2$, from bottom to top.}
	\label{fig:kernel}
\end{figure}

From the previous discussion it follows that the lensing dispersion of the distance modulus can be written as

\begin{eqnarray}
\sigma^2_\mu(z)&=&\frac{15\;\mathrm{mag}^2}{8\pi^2}\;\mathrm{Log}^2(e)\frac{H_0^4\Omega_{\mathrm{m},0}^2}{c^4} \int_0^{+\infty}\frac{\mathrm{d}k}{k^2}(1+z)^2\times\nonumber\\
&\times& P(k,z)\left[ k\;c\Delta\eta(z) \right]^3 \int_0^{k\;c\Delta\eta(z)} dx \frac{\sin x}{x}\;.
\end{eqnarray}
In order to better understand how the nonlinear matter power spectrum gets sliced as a function of redshift and scale when studying this observable, we now define and study the integration kernel 

\begin{equation}\label{eqn:kernel}
g(k,z) \equiv (1+z)^2k\;\left[c\Delta\eta(z)\right]^3\int_0^{k\;c\Delta\eta(z)} dx \frac{\sin x}{x}
\end{equation}
(not to be confused with the growth factor of density fluctuations). In Figure \ref{fig:kernel} we show the shape of this integration kernel as a function of scale for a number of different redshifts. As can be seen, the kernel increases in amplitude with increasing redshift, which is expected because the higher the SN Ia redshift the larger the amount of LSS that is transversed by its light. In the subsequent integration this is partially compensated by the growth of the matter power spectrum, so that overall the lensing dispersion grows with redshift. The behavior with spatial frequency is even more interesting, as it is evident that the kernel always tends to suppress large scales and instead to give importance to small scales. More specifically, when $k \gg c\Delta\eta(z)$ the kernel grows linearly with spatial frequency. The immediate consequence of these considerations is that the lensing dispersion of the distance modulus is highly sensitive to the small-scale matter clustering and it is thus ideally suited to probe the nonlinear growth of cosmic structures. It is worth emphasizing the fact that by looking at the lensing dispersion, one is effectively filtering out the linear regime, which is known to high accuracy, and looking directly into the regime that we know less about.

\begin{figure}
	\centering
	\includegraphics[width=\hsize]{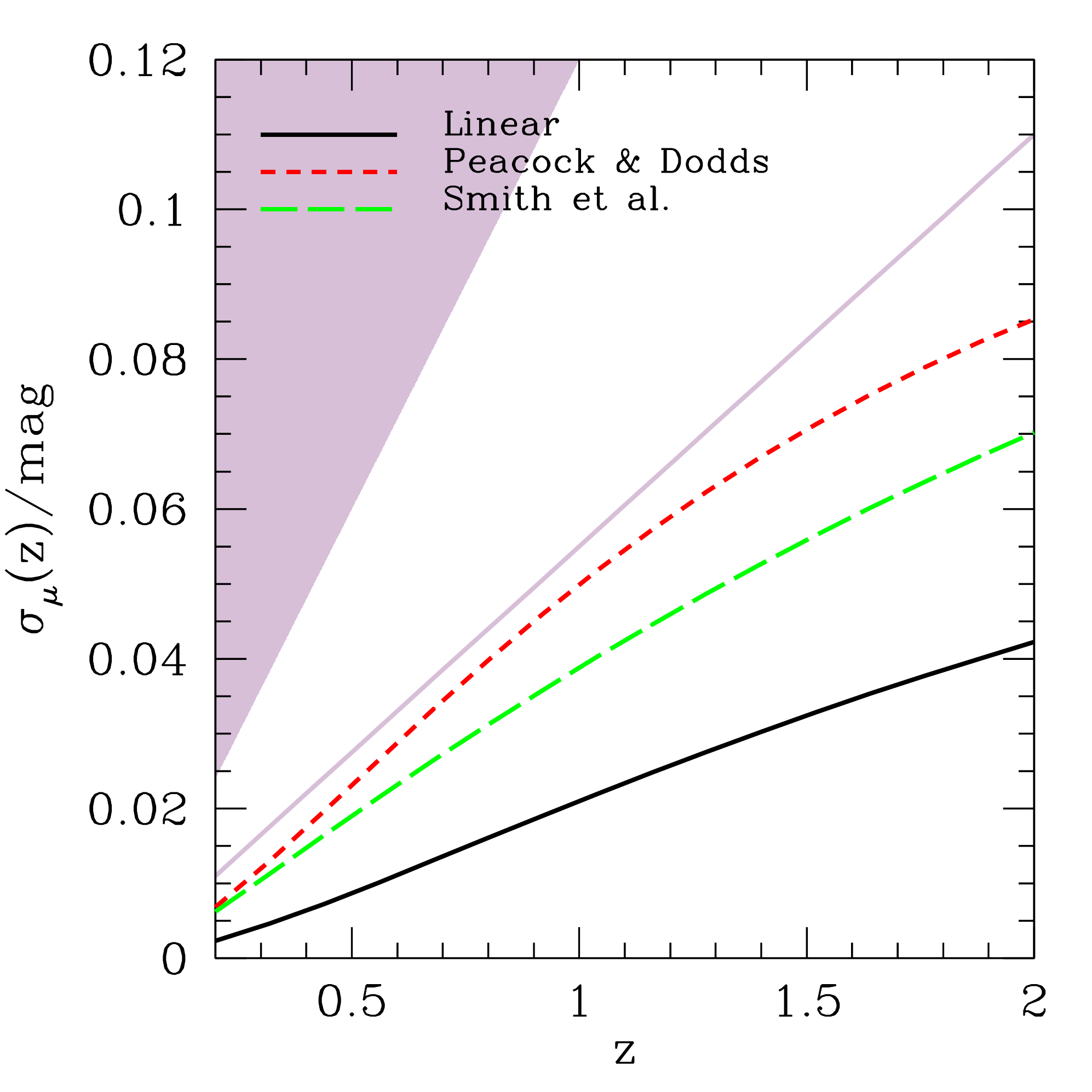}
	\caption{The lensing-induced dispersion of the distance modulus as a function of redshift. Different line types show the results of assuming the linear matter power spectrum (solid black line), the nonlinear power spectrum according to the fit given in \citet{PE94.1} (red short-dashed), and the nonlinear power spectrum according to the fit given in \citet{SM03.1} (green long-dashed). The violet solid line shows the best fit to the lensing dispersion measured for the SNLS sample in \citet{JO10.1}, while the violet band shows the region observationally excluded by the same sample at $95\%$ CL.}
	\label{fig:dispersion}
\end{figure}

In Figure \ref{fig:dispersion}, we make this point more clear by showing the lensing dispersion of the distance modulus as a function of redshift, computed by using different prescriptions for the matter clustering. As can be seen there is a substantial difference between the lensing dispersions induced by linear structures and by nonlinear structures. If one uses the fully nonlinear matter power spectrum rather than its linear counterpart, he or she finds that the dispersion increases by $\sim 80\%$ and up to a factor of $\sim 2.4$ at $z\sim 1$, depending on the prescription that is adopted. Even more interesting is the difference in the results given by the prescription of \citet{PE94.1} for the nonlinear mass clustering and those given by the prescription of \citet{SM03.1}. The lensing dispersion is found to be $\sim 30\%$ higher in the former case with respect to the latter at $z\sim 1$. The extent of this difference between two recipes that are intended to represent the exact same quantity gives an idea of the power of the lensing dispersion to probe the nonlinear structure growth.

In Figure \ref{fig:dispersion}, we also display some observational results. As it has been mentioned in the Introduction, the effect of lensing on the apparent magnitude of high-$z$ SN Ia has been sought for by correlating the residuals in the Hubble diagram with the mass distribution of foreground galaxies. Early results with this approach \citep{WI04.1} found a lensing-induced scatter of $\sim 0.1$ mag for $z\lesssim 0.5$. However, subsequent studies \citep{ME05.2,WA05.1} show that such value was probably an overestimate, and failed themselves to find a conclusive detection. More recently, J\"onsson and collaborators \citep{JO07.1} found a weak correlation for the high-$z$ supernovae in the GOODS field, and later the same authors estimated the lensing-induced scatter in the apparent magnitude of SN Ia as a function of redshift for the Supernova Legacy Survey (SNLS) sample \citep{JO10.1} (see also \citealt{KR10.1}). Specifically, it has been found that the lensing dispersion grows approximately linear with redshift, $\sigma_\mu(z) = Bz$, where the slope is found to be $B \sim 0.06$ mag. This is shown with a violet solid line in Figure \ref{fig:dispersion}. We stress that the observational data do not extend up to $z = 2$ but only up to $z\sim1$, so that the line in the Figure is just an extrapolation based on the power-law assumption. It should also be kept in mind that this measure is affected by rather large uncertainties. J\"onsson and collaborators \citep{JO10.1} found a statistical error on the slope of $\Delta B \sim \pm0.04$ mag at $68.3\%$ Confidence Level (CL), while at $95\%$ CL only an upper limit could be set, $B\lesssim 0.12$ mag. This upper limit is highlighted by the violet shaded area in Figure \ref{fig:dispersion}. To give an idea of the systematic uncertainties involved in these measurements, it should be mentioned that the lensing-induced scatter in the distance modulus of SN Ia is particularly sensitive to the high magnification tail of the magnification distribution (see also \citealt*{MA13.1,QU13.1}), so that samples of even hundreds of SN Ia can lead to a substantially underestimated value of $B$. Kronborg and collaborators \citep{KR10.1} performed an analysis similar to the one of \citet{JO10.1} on the same SN Ia sample, finding $B \sim 0.05 - 0.08$ mag (depending on their assumptions), in agreement with J\"onsson and collaborators but with a smaller statistical uncertainty of $\sim\pm 0.02$ mag at $68.3\%$ CL. Numerical ray-tracing simulations are also affected by substantial systematics, although they generically find lensing-induced dispersions of the apparent magnitudes that are in broad agreement with the aforementioned observational estimates. For instance, \citet{HO05.1} reported a value of $B\sim 0.09$ mag, while \citet{BE00.1} stated that $\sigma_\mu(z = 1)\sim 0.04$ mag. 

\begin{figure}
	\centering
	\includegraphics[width=\hsize]{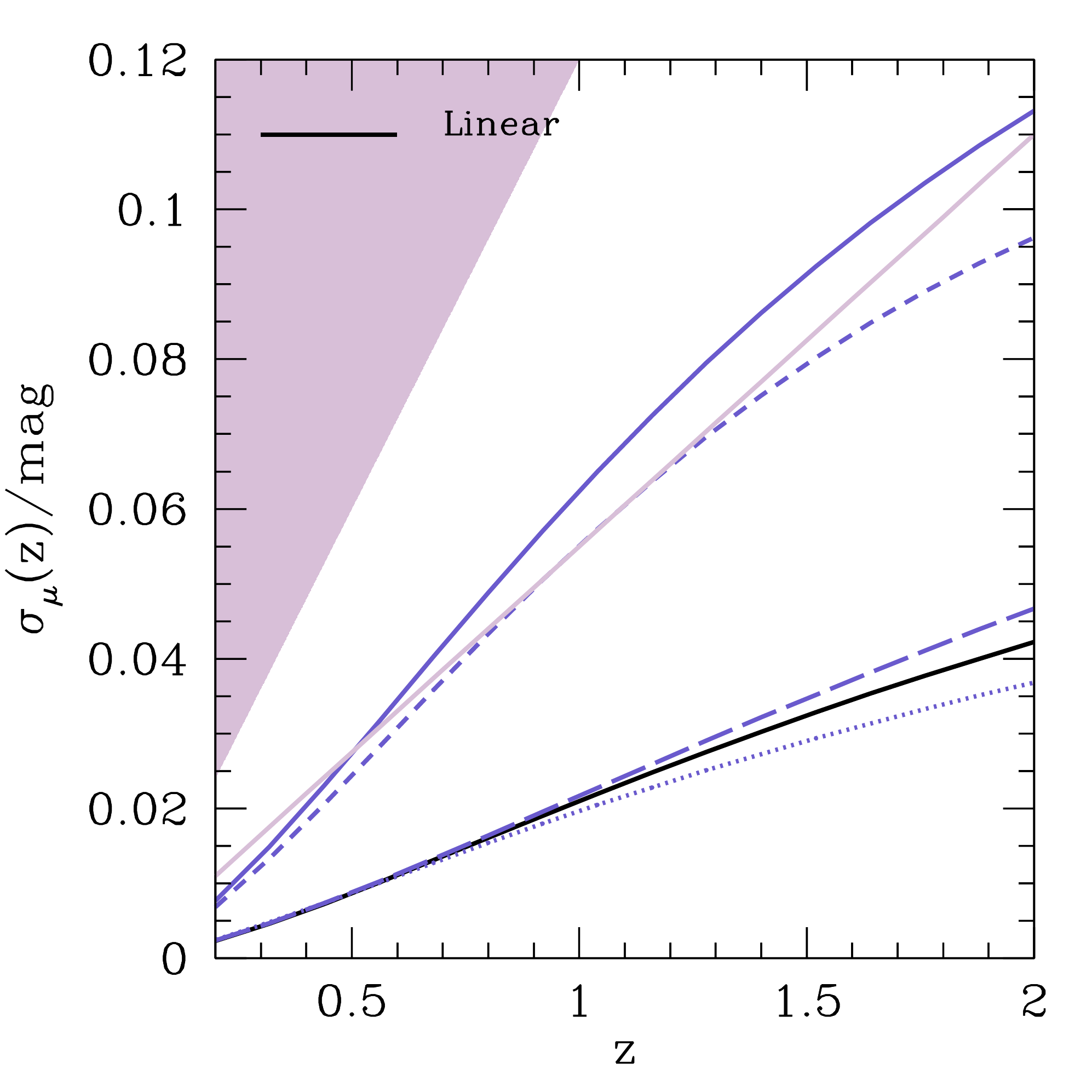}
	\caption{The lensing-induced dispersion of the distance modulus. The black solid line shows the result obtained by using the linear matter power spectrum, while the solid blue line is obtained by modeling the nonlinear matter clustering through the halo model, with fiducial parameter values. The dashed lines show the contributions to the latter of different ranges in scale: $\mathrm{Log}(k\;\mathrm{Mpc}/h)< 0$ (dotted line), $0\le \mathrm{Log}(k\;\mathrm{Mpc}/h)\le 2$ (short-dashed line), and $\mathrm{Log}(k\;\mathrm{Mpc}/h)> 2$ (long-dashed line). The violet solid line shows the best fit to the lensing dispersion measured for the SNLS sample in \citet{JO10.1}, while the violet band shows the region observationally excluded by the same sample at $95\%$ CL.}
	\label{fig:dispersion_dK}
\end{figure}

\begin{figure}
	\centering
	\includegraphics[width=\hsize]{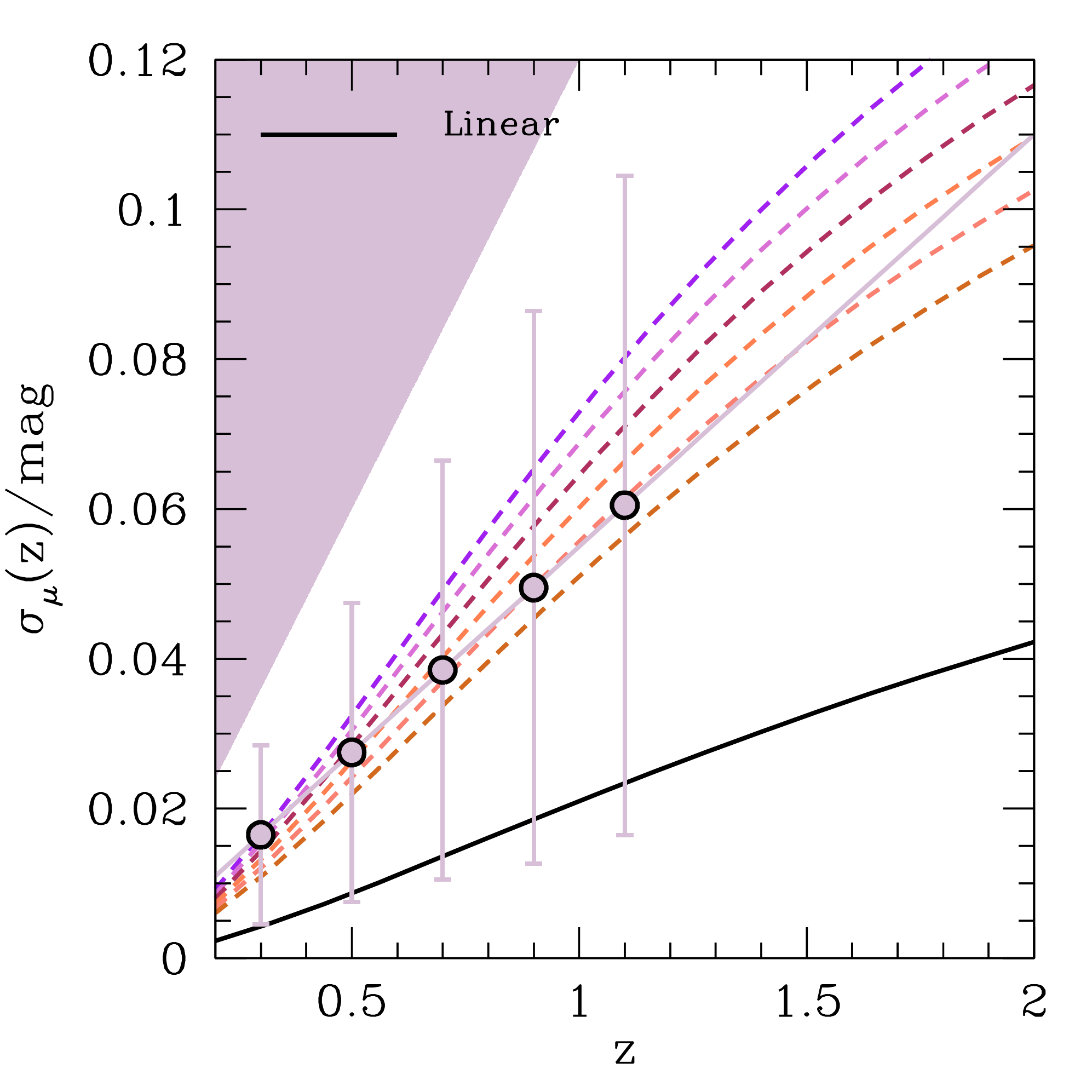}
	\caption{The lensing-induced dispersion of the distance modulus as a function of redshift. The solid black line shows the results of assuming the linear matter power spectrum. The colored dashed lines are computed by modeling the nonlinear matter power spectrum through the halo model, and changing the average halo concentration at all masses (from $0.5$ to $1.5$ times the fiducial value, from bottom to top, in steps of $0.2$). The violet solid line shows the best fit to the lensing-induced scatter measured for the SNLS sample in \citet{JO10.1}, while the violet band shows the region observationally excluded by the same sample at $95\%$ CL. The circles are five mock measurements assumed to lie on the best fit relation and with errorbars corresponding to the $68.3\%$ uncertainty from \citet{JO10.1}.}
	\label{fig:dispersion_HM}
\end{figure}

\section{Modeling the nonlinear matter power spectrum}\label{sct:spectrum}

As a next step, we modeled the nonlinear matter power spectrum by using the halo model \citep{MA00.3,SE00.1,CO02.2}. The halo model is a semi-analytic framework for estimating the mass clustering. It is based on the assumption that all matter in the Universe is locked into gravitationally bound structures (halos), and returns the nonlinear power spectrum as the sum of two terms: the $1-$halo contribution dominates at small scales (the most important in this case), and it depends on the average density profile of halos as a function of mass; the $2-$halo contribution dominates at large scales and depends on the mutual clustering of halo centers. 

In our implementation of the halo model, the mass function has been represented by the \citet{TI08.1} prescription, while for the linear halo bias we adopted the \citet{TI10.1} recipe (see \citealt{FE14.1} for additional details). Furthermore, we assumed the average halo density run to be a NFW profile \citep*{NA96.1}, and correlated its virial mass to its concentration parameter via the following relation, which is in good agreement with the numerical $n-$body results of Duffy and collaborators \citep{DU08.1}:

\begin{equation}\label{eqn:cm}
c_\mathrm{F}(m,z) = \frac{11}{1+z} \left( \frac{m}{10^{12}h^{-1}M_\odot} \right)^{-0.1}\;.
\end{equation}
The suffix F in $c_\mathrm{F}$ means that this is our fiducial concentration-mass relation. The solid blue line in Figure \ref{fig:dispersion_dK} shows the lensing dispersion as a function of redshift computed using the halo model with this fiducial choice of parameters.

As can be seen the dispersion produced by the halo model is substantially larger than that produced by the fits to numerical $n-$body simulations (up to $\sim 40\%$ at $z\sim 1$). This can be traced back again to the sensitivity of this probe to small scales, and to the fact that fits are accurate only up to a few tens $h$ Mpc$^{-1}$ for $z > 0$. Indeed, the prescriptions of \citet{PE94.1} and \citet{SM03.1} are known to substantially underestimate the matter clustering power for $k\gtrsim 10h$ Mpc$^{-1}$ \citep{HI09.1} (see also \citealt{TA12.1} for an improvement over these prescriptions). In Figure \ref{fig:dispersion_dK} we also give a further characterization of the dependence of lensing dispersion on small-scale mass clustering. Specifically, we display the contribution of different scale ranges to the total lensing dispersion as a function of redshift. As can be seen, the bulk of the signal comes from nonlinear scales, $0\le \mathrm{Log}(k\;\mathrm{Mpc}/h)\le 2$, which give $\sim 90\%$ of the total dispersion at $z\sim1$. The remaining $\sim 10\%$ is distributed almost equally amongst smaller and larger scales, which both give a contribution comparable to the linear calculation. This can be seen as a further confirmation of the sensitivity of the lensing-induced dispersion to the nonlinear structure formation. A real-space analog of Figure \ref{fig:dispersion_dK} can be found in \cite{KA11.1}.

The implementation of the halo model that we adopted here does not account for the presence of substructures within halos, nor the stochastic distribution of their concentrations for a given mass. This means that the prescription of Eq. (\ref{eqn:cm}) probably underestimates the clustering power at small scales \citep{GI10.1}. It should also be kept in mind that baryonic physics can have a substantial impact at those scales. If one wants to keep representing virialized structures as NFW halos even when baryons are present (while this is not extremely accurate, it would suffice for a preliminary and illustrative study), then one should modify the concentration-mass relation in order to account for the redistribution of dark matter caused by gas physics \emph{and} the distribution of baryons themselves. If this is the case, then gas cooling and star formation would increase the value of $c_\mathrm{F}$, while a strong feedback from supernovae and/or energy injection from active galactic nuclei (AGN) would decrease it, even substantially \citep{SE11.1,ZE13.1}.

In order to illustrate these effects, we modified the fiducial concentration-mass relation by defining a new concentration $c(m,z)$ such that, if $p\equiv c/c_\mathrm{F}$, then

\begin{equation}
p(m,z) = p_0\left( \frac{m}{10^{12}h^{-1}M_\odot} \right)^\alpha\;.
\end{equation}
In Figure \ref{fig:dispersion_HM} we show the dispersion of the distance modulus resulting by the assumption of $\alpha = 0$ (i.e., the same mass dependence of the fiducial relation) and a number of different values for $p_0$, both smaller and larger than unity. Evidently, by changing the concentrations of virialized structures we obtain a substantial modification of the lensing dispersion, especially at high redshift. This suggests that observations of the lensing dispersion can be used to put constraints on the internal structure of halos, and thus eventually on the physical processes (baryonic and not) at work there. In the next Section, we explore this venue by using as mock observational data the five points that are shown in Figure \ref{fig:dispersion_HM}. These have been placed exactly on the best fit line provided in \citep{JO10.1}, with errors corresponding to the $68.3\%$ confidence intervals given in the same work.

\section{Constraints on nonlinear structure growth}\label{sct:constraints}

In order to exemplify the power of the lensing-induced dispersion in the apparent brightness of SN Ia for gathering information on the nonlinear growth of structures, we considered five observational points placed along the straight line $\sigma_\mu(z) = (0.055\pm 0.040)\;z$ mag (the best fit reported in \cite{JO10.1} with $68.3\%$ CL errors) at redshifts between $0.3$ and $1.1$ equally spaced by $0.2$. These points and the respective error bars are shown in Figure \ref{fig:dispersion_HM}. As the statistical uncertainty on the redshift slope is symmetric\footnote{The $68.3\%$ uncertainties reported by \cite{JO10.1} are actually $\Delta B = ^{+0.039}_{-0.041}$ mag. As they are very close to symmetric we just assumed $\Delta B = \pm 0.040$ mag throughout.} we assumed the errors on each mock observational point to be Gaussian and thus estimated a standard $\chi^2(p_0,\alpha)$ function for the two parameters of the mass-concentration relation in the halo model. Obviously, this procedure is highly idealized. The mock observational points do not come from actual observations (although they agree with actual observations), and we have no way to assess the covariance between different points, which is certainly present. Yet, the error on individual observations would likely be less than the uncertainty considered here (see Figure 7 of \citealt{JO10.1}), which should at least partly compensate for ignoring covariances. More in general, we believe this approach to be sufficient for illustrative purposes. Note that there exist other and likely more precise methods to probe the inner structure of galaxies and galaxy clusters. However, the SN Ia lensing dispersion is sensitive to different scales and different mass ranges than those other methods, thus it returns at least complementary information. In any case, here we just mean to give a demonstration of one possible application of SN Ia lensing dispersion, and discuss more applications in Section \ref{sct:conclusions}.

\begin{figure}
	\centering
	\includegraphics[width=\hsize]{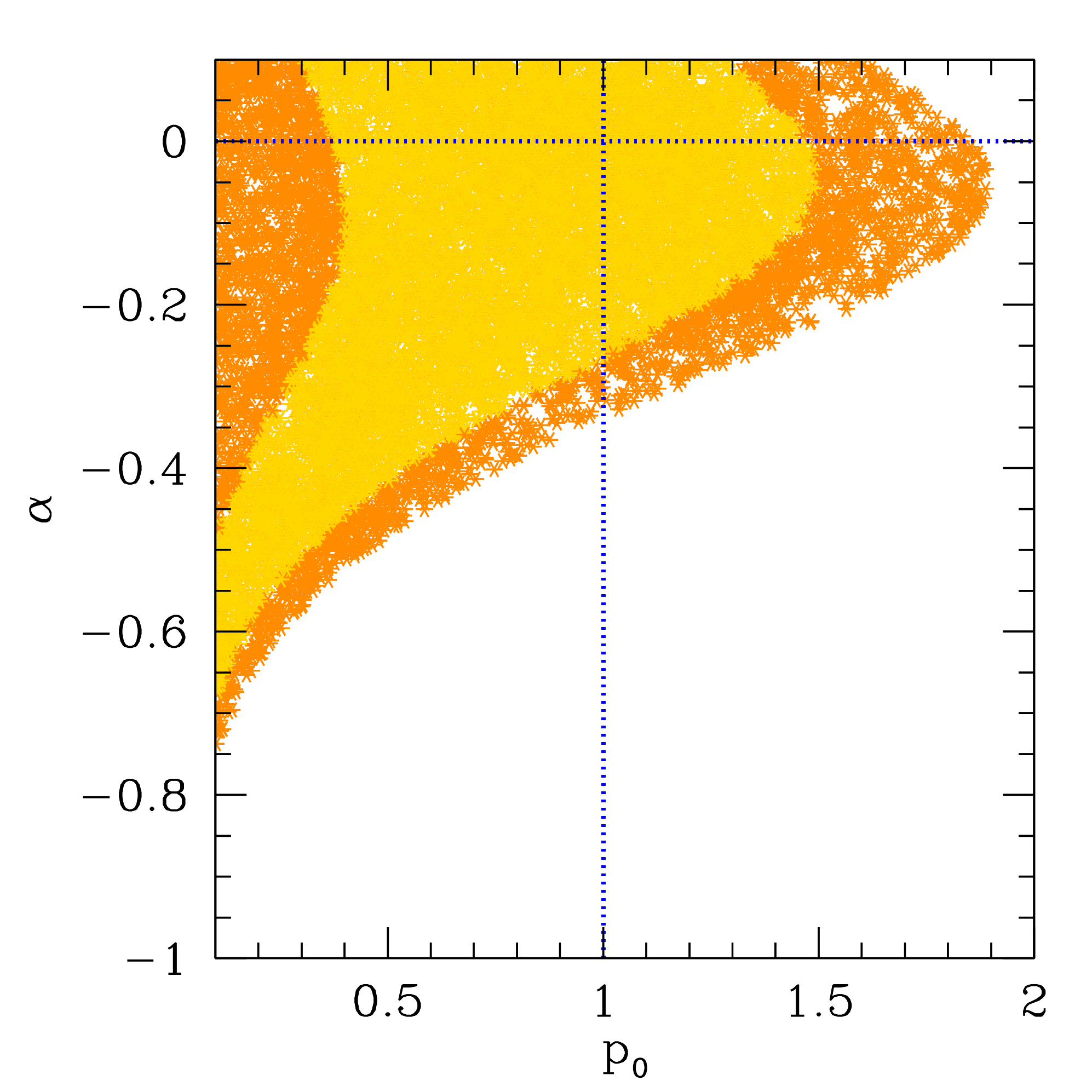}
	\caption{Confidence regions in the $(p_0,\alpha)$ plane, as sampled by MCMCs. The yellow points show the $68.3\%$ confidence area, while the orange points highlight the $95.5\%$ confidence region. The intersection of the two blue dotted lines represent the fiducial model given by $p_0=1$ and $\alpha=0$.}
	\label{fig:mcmc}
\end{figure}

As can be inferred by inspecting Figure \ref{fig:dispersion_HM}, current observational uncertainties do not allow to put strong constraints on halo concentrations. To confirm this, we explored parameter space in search for the maximum likelihood via Monte Carlo Markov Chains (MCMCs) implemented through a Metropolis-Hastings algorithm. Specifically, we ran three MCMCs with different and randomly chosen starting points in parameter space, in order to make sure not to miss any secondary maximum. Each chain was ran for $5,000$ steps with an initial burn-in of $200$ steps that were discarded. In Figure \ref{fig:mcmc} we show the CLs in the parameter plane given by $p_0$ and $\alpha$. We decided to adopt a flat prior for $p_0>0.1$, as lower concentrations would be unrealistically small, and for $\alpha<0.1$, as we wanted concentrations to still be decreasing with mass. The latter requirement is motivated by the fact that observations of galaxy groups and clusters do find lower concentrations for higher mass systems \citep{BU07.1,SC07.2,OG09.1}, thus suggesting that baryons do not alter the sign of the slope of the concentration-mass relation.

As can be seen in Figure \ref{fig:mcmc}, the constraints on the concentration-mass relation given by SN Ia lensing dispersion are rather loose, confirming the findings presented in Figure \ref{fig:dispersion_HM}. Nevertheless, we are able to exclude a substantial part of the parameter space even with the currently limited observations. Specifically, the slope $\alpha$ of the concentration ratio-mass relation is not allowed to be too negative, especially when the normalization $p_0$ is larger than unity: values of $\alpha \lesssim -0.3$ for $p_0 > 1$ are excluded at more than $95.5\%$ CL. Another interesting fact that can be inferred from the Figure is that values of $p_0<1$ are slightly preferred with respect to $p_0>1$. Although this cannot yet be conclusive, it hints at structures being on average less compact than suggested by numerical $n-$body simulations. We shall elaborate on this further down in Section \ref{sct:conclusions}. Given the shape of the posterior probability distribution, we are only able to put upper bounds on $p_0$ and lower bounds on $\alpha$. Specifically, by marginalizing the posterior over the other parameter, we obtain that the normalization of the concentration ratio-mass relation is constrained as $p_0 < 1.00$ at $68.3\%$ CL and $p_0 < 1.64$ at $95.5\%$ CL, while the slope is constrained as $\alpha > -0.26$ at $68.3\%$ CL and $\alpha > -0.52$ at $95.5\%$ CL.

\begin{figure}
	\centering
	\includegraphics[width=\hsize]{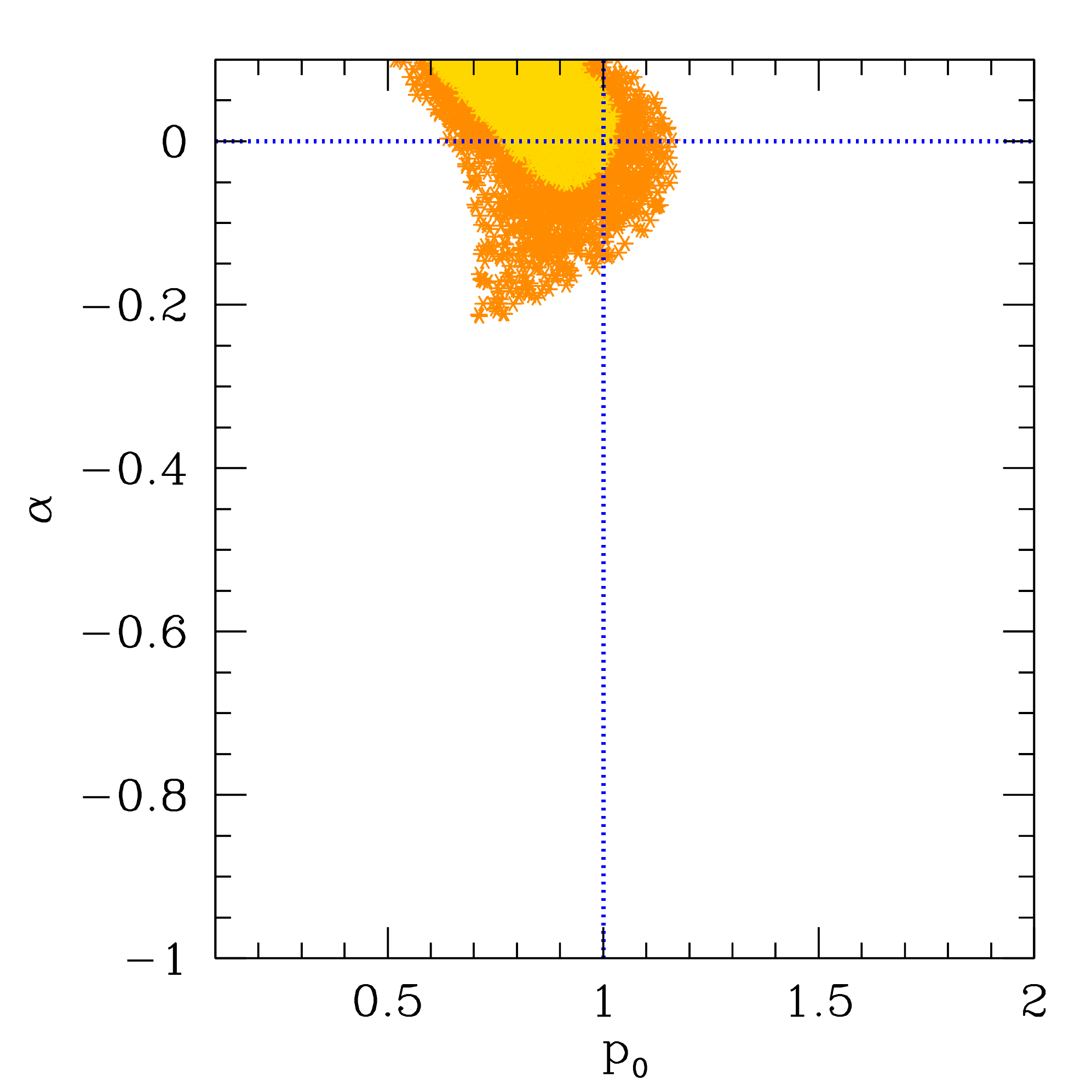}
	\caption{The same as Figure \ref{fig:mcmc} obtained after reducing the statistical errors on the mock observations (see Figure \ref{fig:dispersion_HM}) by a factor of $5$, in order to mimic future experimental improvements.}
	\label{fig:mcmc_REDUCED_ERRORS}
\end{figure}

In order to take into account future improvements on the statistics of SN Ia, we repeated the calculations described above after reducing the errors on the mock observational points of Figure \ref{fig:dispersion_HM} by a factor of $5$. This is consistent with the improvement in statistics that will be provided by \emph{Euclid}, and it is probably a conservative estimate given the number of supernovae that will be available after the Large Synoptic Survey Telescope (LSST) survey (see the discussion in Section \ref{sct:conclusions} below) and the likely advancement in modeling expected in forthcoming years\footnote{Different surveys actually cover different redshift ranges, however here we are interested only into conveying the general idea. Thus, we decided to ignore these differences and used the same mock observational points as before}. The likelihood regions on $p_0$ and $\alpha$ explored by the MCMCs under this assumption are shown in Figure \ref{fig:mcmc_REDUCED_ERRORS}. As can be seen, in this case we obtain remarkable constraints on the concentration-mass relation of cosmic structures, with only a very small region of parameter space being allowed. By marginalizing the posterior probability distribution over the other parameter we find that the normalization is constrained as $0.65 < p_0 < 0.91$ at $68.3\%$ CL and $0.50 < p_0 < 1.10$ at $95.5\%$ CL, while the slope is constrained as $\alpha > 0.00$ at $68.3\%$ CL and $\alpha > -0.14$ at $95.5\%$ CL. In the next Section \ref{sct:conclusions} we discuss the implications of these bounds for the internal structure of dark matter halos.

Before concluding this Section we would like to stress a few points. First, for a more complete treatment of this problem we could have added a third parameter to the model, modulating the change in redshift of the halo concentration. This would have been highly degenerate with $p_0$, and given the weakness of observational data we preferred to opt for simplicity. Second, and as mentioned in the Introduction, we could also have chosen a slightly different approach, namely fit the total measured dispersion in SN Ia apparent magnitude (including lensing and intrinsic contributions) and marginalize over the intrinsic part of the dispersion itself after modeling it (see \citealt*{DO06.2,MA13.1,QU13.1}). Again, this approach would be more suitable when more secure data become available. In the future, the increased statistics will allow one to improve upon both these points.

\section{Discussion and conclusions}\label{sct:conclusions}

In this paper, we investigated the effect of nonlinear matter clustering on the dispersion in the apparent magnitude of SN Ia induced by stochastic lensing fluctuations. As it turns out, this lensing dispersion filters out the linear structures, while highlighting small-scale structure growth. As such, it is an excellent observational probe of the poorly understood nonlinear matter power spectrum. The following results can be summarized:

\begin{itemize}
\item The main contribution to the lensing dispersion is given by structures on sub-Mpc spatial scales, $0 \le \mathrm{Log}(k\;\mathrm{Mpc}/h)\le 2$, constituting $\sim 90\%$ of the total at $z\sim 1$. This is a regime where the mass clustering is highly sensitive to the internal composition of dark matter halos and the physics of baryons.
\item As a consequence of the above, the choice of how to model the nonlinear mass power spectrum has a substantial impact on the theoretical representation of lensing dispersion of SN Ia. Different prescriptions easily produce differences of $\sim 40-50\%$ on the final result.
\item By adopting a halo model for the nonlinear power spectrum we can use the (still marginal) observations of the lensing dispersion to constrain the internal structure of virialized objects. Current statistical errors are still too loose to obtain good constraints, however we find a slight preference for the concentration-mass relation to be somewhat lower than the values found by $n-$body simulations, while we are able to put a lower limit on the slope. 
\item If we artificially reduce the observational error by a factor of $5$, in order to take into account future improvements in statistics and modeling (see below), we find very strong constraints on the concentration-mass relation. For instance, the normalization of the relation is bound to within a very small interval at $68.3\%$ CL, of width $\sim 0.25$.
\end{itemize}

This last point is particularly important, and needs some contemplation. The very tight constraints on the concentration-mass relation that the SN Ia lensing dispersion will be able to provide in the future are going to give important clues on baryonic physics at work during the assembly of structures. Indeed, numerical simulations show that these physical processes are very well capable of robustly redistribute matter (both dark and luminous) within a structure (e.g., \citealt{MC10.1,MC11.1}). In this context, the work by Zentner and collaborators \citep{ZE13.1} is illuminating. The authors of that paper also employed the halo model in order to compute the cosmic shear power spectrum, and then changed the model parameters (that in their case also include a redshift modulation of the concentrations) in order to fit the results of hydrodynamic simulations with different baryonic physical processes \citep{SC10.1}. By moving from an implementation featuring strong AGN winds to ones with negligible feedback of any kind they find that the best fit normalization of the concentration-mass relation can be moved by a factor ranging in a very substantial interval, from $\sim 0.3$ up to $\sim 1.2-1.4$. These numbers cannot be compared directly with our Figure \ref{fig:mcmc} because we do not have any redshift modulation, and our pivot mass for the concentration-mass relation is different from theirs. Yet, they provide a good indication that the lensing dispersion will be very well able to substantially reduce the space of allowed baryonic physical models.

The measurements by \citet{JO10.1} were based on less than $200$ SN Ia. The Union$2.1$ compilation \citep{SU12.1} contains already $\sim 3$ times as many objects (although not all coming from a single homogeneous dataset). In the future, \emph{Euclid} will be capable of observing the light curves of $\sim 3,000$ SN Ia \citep{LA11.2}, while the LSST will increase this number to a staggering $\sim 100,000$ objects \citep{LS09.1}. We can thus expect the measurements of the lensing-induced scatter in the distance modulus of standard candles to become much more precise in forthcoming years (even significantly more than we assumed in our last bullet), and thus this method to emerge as one of the principal probes of nonlinear structures. With increasing statistics the approach that has been exemplified in the present paper can be improved to probe substructure and stochasticity of dark matter halos (see \citealt{GI10.1}) as well as, directly, the effects of gas physics and star formation (thanks to a generalized halo model that our group is developing, see \citealt{FE14.1}).

The previous discussion does not take into account the presence of unavoidable systematics in the measurements of the SN Ia lensing flux dispersion. It is likely that these systematics will not be reducible below a certain (still unknown) level, which is going to obstruct to some extent the constraining power of SN Ia for the nonlinear matter clustering. The presence of this systematics "floor" will need to be addressed properly when the statistics will become large.

The results of this work show not only how the lensing-induced dispersion of apparent magnitudes of SN Ia can lead to interesting constraints on nonlinear structure growth, but also how, and because of this very fact, constraints on cosmological parameters based on this method \citep*{DO06.2,MA13.1,QU13.1,BE13.3} should be taken with grain of salt. Indeed, considering how sensitive this approach is to nonlinear matter clustering, in the future it will be likely a crucial instrument to study how galaxy formation physics contribute to shaping the LSS of the Universe.

\section*{Acknowledgments}

We are grateful to F. J. Nugier and R. B. Metcalf for interesting comments that allowed us to improve the presentation of this work. CF has received funding from the European Commission Seventh Framework Programme (FP7/2007-2013) under grant agreement n$^\circ$ 267251. LM acknowledges financial contributions from contract ASI/INAF I/023/12/0, from PRIN MIUR 2010-2011 "The dark Universe and the cosmic evolution of baryons: from current surveys to Euclid" and from PRIN INAF 2012 "The Universe in the box: multiscale simulations of cosmic structure". We wish to thank an anonymous referee for insightful remarks on the manuscript.

{
\bibliographystyle{mn2e}
\bibliography{./master}
}

\end{document}